\documentclass[a4paper, 11pt,margin=1in]{article}
\usepackage{parskip} 
\usepackage{tikz} 
\usepackage{amsmath}
\usepackage{amssymb}
\usepackage{hyperref}
\usepackage{listings}
\usepackage{dsfont}
\usepackage{graphicx}
\usepackage{subcaption}
\usepackage{multirow}
\usepackage[square,numbers]{natbib} 
\usepackage{float}
\usepackage[ruled,vlined]{algorithm2e}
\usepackage{epstopdf}
\usepackage{authblk}
\usepackage{lineno}

\usepackage{enumitem}
\usepackage{authblk}
\usepackage[margin=1in,includehead,includefoot]{geometry}
\title{A sampling algorithm to compute the set of feasible solutions for non-negative matrix factorization with an arbitrary rank } 

\author{Ragnhild Laursen\thanks{Email: ragnhild@math.au.dk}  \ and Asger Hobolth\thanks{Email: asger@math.au.dk}}
\affil{Department of Mathematics, Aarhus University}
\date{\today}

\begin{document}

\maketitle
\sloppy
\begin{abstract}
    Non-negative Matrix Factorization (NMF) is a useful method to extract features from multivariate data, but an important and sometimes neglected concern is that NMF can result in non-unique solutions. Often, there exist a Set of Feasible Solutions (SFS), which makes it more difficult to interpret the factorization. This problem is especially ignored in cancer genomics, where NMF is used to infer information about the mutational processes present in the evolution of cancer. In this paper the extent of non-uniqueness is investigated for two mutational counts data, and a new sampling algorithm, that can find the SFS, is introduced. Our sampling algorithm is easy to implement and applies to an arbitrary rank of NMF. This is in contrast to state of the art, where the NMF rank must be smaller than or equal to four. For lower ranks we show that our algorithm performs similarly to the polygon inflation algorithm that is developed in relations to chemometrics. Furthermore, we show how the size of the SFS can have a high influence on the appearing variability of a solution. Our sampling algorithm is implemented in an R package \textbf{SFS} (\url{https://github.com/ragnhildlaursen/SFS}).  
    
\end{abstract}

\section{Introduction}
The applications of Non-negative Matrix Factorization (NMF) are many, and one of them is in pure component analysis for analytical chemistry. In this field, it is a big obstacle that the solution from NMF is non-unique, such that there exist a Set of Feasible Solutions (SFS) and not only one. As a consequence, there exist a vast literature in chemometrics on finding the SFS, both for general applications of NMF and more specific applications to chemical data \cite{Borgen1985,Gemper1999,lawton1971,Sawall2013}. Having a non-unique solution to NMF makes it problematic to interpret the factorization. NMF is an unsupervised learning method that factorizes a non-negative data matrix $M \in \mathbb{R}_+^{K \times G}$ into two non-negative matrices $P\in \mathbb{R}_+^{K \times N}$ and $E \in \mathbb{R}_+^{N \times G}$ \cite{lee2001}. The rank $N$ is usually chosen much smaller than $K$ and $G$. This means the factorization is an approximation
\[ M \approx PE.\]
 Here, all the entries in $P$ and $E$ are free parameters, that need to be estimated. The problem with the factorization is that other solutions can be constructed by finding invertible matrices $A \in \mathbb{R}^{N \times N}$ that fulfill 
\begin{equation}
    \tilde{P} = PA \geq 0 \quad \tilde{E} = A^{-1}E \geq 0.
    \label{eq:ineq}
\end{equation}
Then the product of $\tilde{P}$ and $\tilde{E}$ gives the exact same approximation of $M$ as the product of $P$ and $E$. Trivial ambiguities exist when $A$ is either a diagonal matrix, which scales the entries in $P$ and $E$ or a permutation matrix, which re-orders the columns in $P$ and rows in $E$. These ambiguities are always possible, so it is standard to define $P$ and $E$ as a unique solution to NMF if the only possible ambiguities are scaling and/or re-ordering \citep{Stodden2004,Huang2013,Laurberg2008,Brie2005}. Here, the SFS for $P$ and $E$ are defined as 
\begin{align}
    \begin{split}
         \mathcal{M}(P) &= \left\{\tilde{P} \in \mathbb{R}_+^{K \times N} \ \middle| \ \exists \text{ invertible }A \in \mathbb{R}^{N \times N} : \tilde{P} = PA \geq 0 \text{ and } \tilde{E} = A^{-1} E \geq 0 \right\} \\
         \mathcal{M}(E) &= \left\{\tilde{E} \in \mathbb{R}_+^{N \times G} \ \middle| \ \exists \text{ invertible }A \in \mathbb{R}^{N \times N} : \tilde{P} = PA \geq 0 \text{ and } \tilde{E} = A^{-1} E \geq 0 \right\}
    \end{split}
        \label{eq:setPE}
\end{align}
where $A$ is normalised such that the columns of $\tilde{P} = PA$ sum to one, as this removes the scaling ambiguity. This can be shown as follows
\[\sum_{j=1}^K \tilde{P}_{jn} = \sum_{j=1}^K \left\{ \sum_{i=1}^N P_{ji}A_{in}  \right\} =  \sum_{i=1}^N A_{in}  \underbrace{\sum_{j=1}^K  P_{ji}}_{=1} = \sum_{i=1}^N A_{in}.\]
 for each $n = 1, \dots ,N$ and therefore constructing $A$ such that the columns sum to one would automatically assure the columns of $\tilde{P}$ sum to one, such that the whole SFS, $(\tilde{P}, \tilde{E})$, have the same scaling. The problem of re-ordering of the entries is solved by ordering by cosine similarity. 

In chemometrics, the SFS for a given NMF solution has been investigated since 1971, where Lawton and Sylvester \cite{lawton1971} first introduced the analytical calculation for finding the SFS for rank two i.e. $N = 2$. A vast literature has later evolved on finding the SFS for a higher rank, $N \geq 3$, both by analytical calculations and with numerical algorithms \cite{Borgen1985,Gemper1999,HENRY1990,Sawall2017} . The analytical calculations of the SFS have been extended to ranks of three and four \cite{Borgen1985, HENRY1990}, but require a significant amount of calculations when the size of $K$ and $G$ are large, as it increases the number of inequalities in \eqref{eq:ineq}. For a higher rank others have proposed numerical algorithms to approximate the SFS \cite{golshan2016review,Sawall2013}. Here, we will focus on the recent polygon inflation algorithm \cite{Sawall2013} to compare with our new sampling algorithm. For a further review and comparison of some of the other currently available algorithms we refer to \cite{golshan2016review}. Our sampling algorithm has the advantage of being able to compute the SFS for an arbitrary rank of the factorization, which is needed for many NMF applications. In particular we will focus on applications in cancer genomics, where NMF is used to infer information about the mutational processes in cancer evolution. 

In the field of cancer genomics, the assumption is that all the mutations in a cancer genome comes from a spectrum of $N$ mutational process, that each have their own effect on the cancer genome. The overall goal is to identify these mutational processes and use it to enhance the understanding of cancer evolution \cite{nik2017}. A popular method to infer information on the mutational processes from a matrix of mutational counts is NMF. Here, $M$ is a $K \times G$ matrix of mutational counts. The $G$ columns of dimension $K \times 1$ represent the mutational catalog for each genome, which are found from sequencing a cancer genome. The number of different mutation types, $K$, in the data sets is $96$ as they include the $6$ base substitutions C$>$A, C$>$G, C$>$T, T$>$A, T$>$C, T$>$G, and the immediate 3' and 5' neighboring bases, i.e. $6 \cdot 4 \cdot 4 = 96$ different mutation types. To recover the mutational processes from the mutational catalogs, we assume that each mutational catalog is a positive linear combination of a certain number of mutational processes, $N$. This means approximating $M$ by the factorization of two non-negative matrices $P$ and $E$ of dimension $K \times N$ and $N \times G$, respectively, such that $M \approx PE$. The matrix $P$ represents the $N$ signatures for the mutational processes and the matrix $E$ represents their exposures for each mutational catalog. The size of $N$ is usually chosen magnitudes smaller than $K$ and $G$, which makes $P$ and $E$ a low-dimensional representation of $M$ describing its main features.

\begin{figure}[t]
	\centering
	\includegraphics[width = \textwidth]{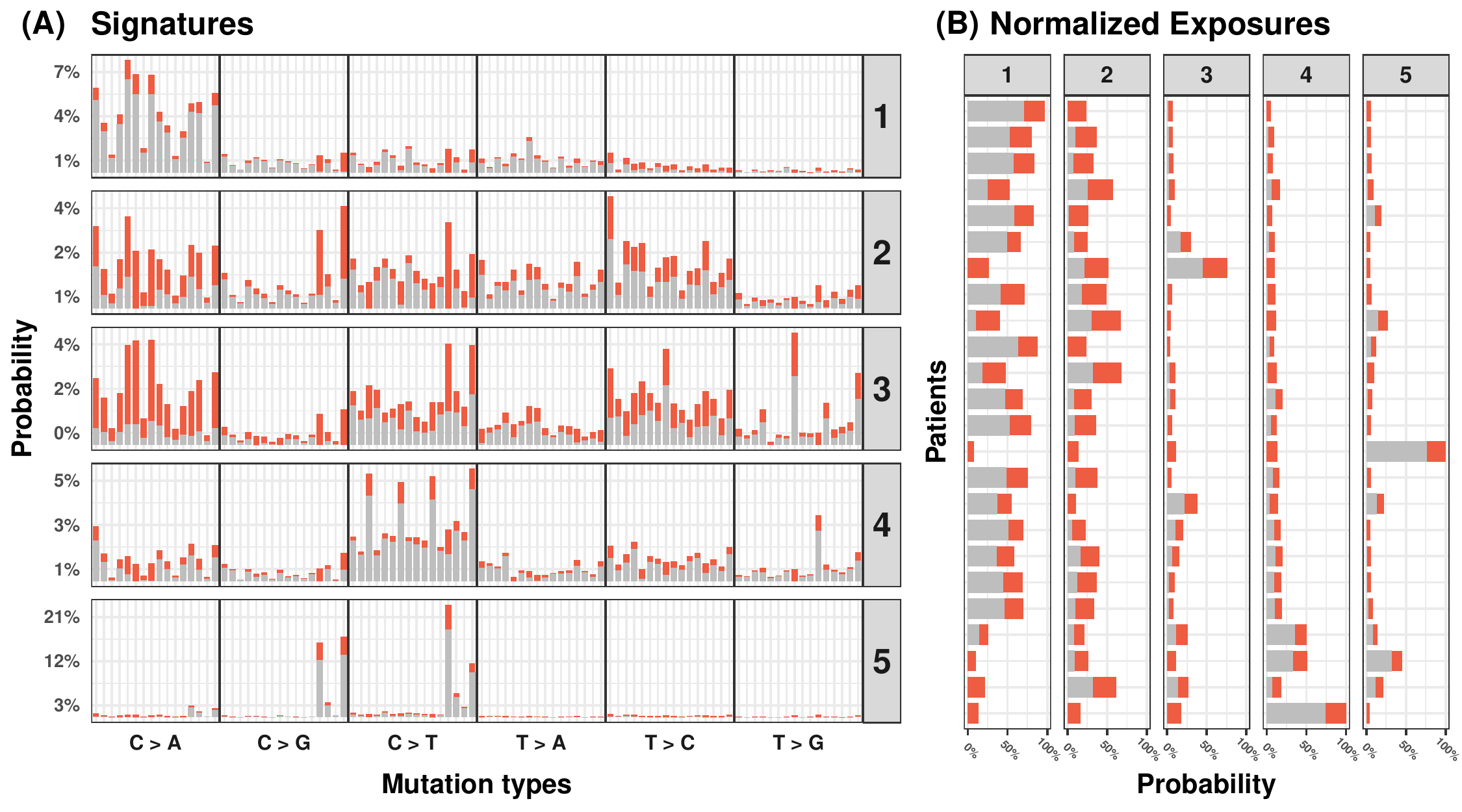}
	\caption{The signatures and exposures for the Lung A. cancer data, when assuming five mutational processes. In \textbf{(A)} the signatures are depicted, with the SFS marked in red from the minimum feasible value to the maximum feasible value. In \textbf{(B)} the normalised exposures are depicted, where the SFS is again marked in red. The signatures are arranged according to their total exposure i.e. $\sum_{g=1}^G E_{ng}$ in decreasing order. }
	\label{fig:LACAsfs}
\end{figure}

Through this paper we refer to two different data sets of mutations in whole genomes. One of them is the 21 Breast cancer genomes, which is a commonly known and analysed data set in relation to mutational processes in cancer genomics \cite{alex2013,fischer2013,signeR2017}, which we refer to as the Breast cancer data. The data is a matrix of dimension $96 \times 21$ consisting of the mutational counts for the $96$ different mutationtypes in $21$ different breast cancer genomes. The other data set consists of the mutational counts from 24 patients having Lung Adenocarcinoma cancer taken from \cite{alex2013signatures}. The data set is referred to as the Lung A. cancer data and is a $96 \times 24$ matrix of mutational counts. The Lung A. cancer data is chosen because it has a large SFS, even for large sizes of $N$. The SFS found with our sampling algorithm for the Lung A. cancer when assuming five mutational processes is depicted in Figure \ref{fig:LACAsfs}. The red bars illustrate the SFS from the minimum feasible value to the maximum feasible value for each of the different entries. In Figure \ref{fig:LACAsfs} we observe that the range of the SFS varies over the different mutation types and patients, as the changes are very dependent on the structure of one another. The code and data for this paper is available at \url{https://github.com/ragnhildlaursen/sampleSFS_paper}.

The paper is structured as follows. In Section 2 we introduce and explain our sampling algorithm. In Section 3 we apply our algorithm on the Breast and Lung A. cancer data and compare it to the polygon inflation algorithm. Section 4 is a further analysis of the optimal choice of parameters as well as running time of our sampling algorithm. In Section 5, we at last cover the problem of identifiability of $P$ and $E$ in relation to initialization for the updates made by Lee and Seung \cite{lee2001} and noise in the data. These identifiability problems are compared to the influence of the SFS. The paper ends with concluding remarks regarding our sampling algorithm, choice of rank $N$, and the lack of uniqueness of the matrix factorization. 

\section{The sampling algorithm} \label{sec:sampling}
Recall that our starting point is two non-negative matrices $P\in \mathbb{R}_+^{K \times N}$ and $E \in \mathbb{R}_+^{N \times G}$ that approximates our data $M \in  \mathbb{R}_+^{K \times G}$, and that we want to find the SFS in \eqref{eq:setPE} for both $P$ and $E$. The general idea of our algorithm is to use the simple analytical calculation for rank two, and adapt it to higher dimensions through sampling. First, the SFS is described in the simple setting of $N = 2$ to ease the understanding of the general setting.    
\subsection{SFS for $N = 2$}
For the case of a rank two factorization, the SFS for $P$ and $E$ can be found in a closed form. The calculations we make here are similar to the ones found in \cite{Brie2005}, but here we find the SFS for one column of $P$ at a time. Assume we would like to find the SFS for the first column of $P$. Then we set
\begin{equation}
     A_{12}(\lambda) = \begin{pmatrix}
1-\lambda & 0 \\
\lambda & 1 
\end{pmatrix} .
\label{eq:A2}
\end{equation}
The inverse $A_{12}^{-1}(\lambda)$ is simple as well and can be directly expressed in terms of the original matrix as
\begin{equation}
   A_{12}^{-1}(\lambda) = \frac{1}{1-\lambda} \begin{pmatrix}
1 & 0 \\
-\lambda  & 1 - \lambda
\end{pmatrix} =  \begin{pmatrix}
1 + \frac{\lambda}{1-\lambda} & 0 \\
-\frac{\lambda}{1-\lambda}  & 1
\end{pmatrix} = A_{12} \left( -\frac{\lambda}{1-\lambda} \right).
\label{eq:inverse} 
\end{equation}
The simple inverse eases both the calculations and computation time for the algorithm. The feasible values of $\lambda$ must fulfill that all entries of $\tilde{P} = PA_{12}(\lambda)$ and $\tilde{E} = A_{12}^{-1}(\lambda)E$ remain non-negative, which can be formulated as
\begin{equation}
    0 \leq PA_{12}(\lambda) = \begin{pmatrix}
    P_{11}(1-\lambda) + P_{12} \lambda & P_{12} \\
    \vdots & \vdots \\
    P_{K1}(1-\lambda) + P_{K2} \lambda & P_{K2}
    \end{pmatrix} = \begin{pmatrix}
    P_{11} - \lambda(P_{11} - P_{12}) & P_{12} \\
    \vdots & \vdots \\
    P_{K1} -\lambda(P_{K1} - P_{K2}) & P_{K2}
    \end{pmatrix}
    \label{eq:Pineq}
\end{equation}
and 
\begin{align}
\begin{split}
    0 \leq A_{12}^{-1}(\lambda)E &= \frac{1}{1-\lambda}
    \begin{pmatrix}
    E_{11} & \hdots & E_{1G} \\
    -E_{11} \lambda + E_{21} (1 - \lambda) & \hdots & -E_{1G} \lambda + E_{2G} (1 - \lambda)
    \end{pmatrix} \\
    &= \frac{1}{1-\lambda}
    \begin{pmatrix}
    E_{11} & \hdots & E_{1G} \\
    E_{21} - \lambda (E_{11} + E_{21}) & \hdots & E_{2G} - \lambda (E_{1G} + E_{2G})
    \end{pmatrix}.
    \end{split}
    \label{eq:Eineq}
\end{align}
A general requirement is $\lambda < 1$ to assure the entries in \eqref{eq:Eineq} remain non-negative. The entries in \eqref{eq:Eineq} result in an additional upper bound for $\lambda$ given by
\begin{equation}
    \overline{\Lambda}_{12} = \min_{g = 1, \dots G}\left\{ \frac{E_{2g}}{E_{1g} + E_{2g}} \middle| E_{1g} + E_{2g} > 0 \right\} \geq 0.
    \label{eq:upper}
\end{equation}
In \eqref{eq:Pineq}, the requirement of $\lambda < 1$ assures $P_{k1} -\lambda(P_{k1} - P_{k2}) \geq 0$ for all $k$ where $P_{k1} \geq P_{k2}$. For the other case where $P_{k2} > P_{k1}$ we get the following lower bound for $\lambda$
\begin{equation}
    \underline{\Lambda}_{12} = \max_{k=1,\dots, K} \left\{ \frac{P_{k1}}{P_{k1} - P_{k2}} \middle| P_{k2} > P_{k1} \right\} \leq 0.
    \label{eq:lower}
\end{equation}
This means that all $\lambda \in [\underline{\Lambda}_{12},\overline{\Lambda}_{12}]$ give feasible solutions for the first column of $P$, while the second column is fixed. The bounds of $\lambda$ for the second column of $P$ will be similar, but where $1$ and $2$ are simply switched in \eqref{eq:upper} and \eqref{eq:lower}. In this simple setting, the individual feasible intervals of $[\underline{\Lambda}_{12},\overline{\Lambda}_{12}]$ and $[\underline{\Lambda}_{21},\overline{\Lambda}_{21}]$ give the whole SFS for both $P$ and $E$. In the case of a rank higher than two the analytical calculations for the SFS are substantially more complicated. Calculations for $N$ equal to $3$ can be seen in \cite{Borgen1985,HENRY1990}. 

\subsection{Sampling SFS for arbitrary rank}
Our algorithm uses the analytical calculations for $N = 2$ above and is inspired by Gibbs Sampling \cite{gibbs1990}. The idea is to change each column of $P$ with an affine combination of another column chosen by random sampling. This is done sequentially for each column of $P$ while updating $E$ correspondingly such that the matrix product remains the same. For the approach we define a general transformation matrix $A_{ij}(\lambda)$, of dimension $N \times N$, similar to the one in \eqref{eq:A2}
\begin{equation}
\big( A_{ij}(\lambda) \big)_{uv} = 
 \begin{cases}
1 - \lambda & \text{ if } u = v = i\\
1 & \text{ if } u = v \neq i \\
\lambda & \text{ if } u = j, v = i \\
0 & \text{ otherwise. }
\end{cases}
\label{eq:T}
\end{equation}
The transformation $A_{ij}(\lambda)$ changes column $P_i$ to an affine combination of column $P_i$ and $P_j$, where $j \neq i$. Let $\Lambda_{ij}$ denote the feasible interval for $\lambda$ given a solution $P$ and $E$, such that $PA_{ij}(\lambda) \geq 0$ and $A^{-1}_{ij}(\lambda)E \geq 0$. The endpoints of this interval are similar to the ones in \eqref{eq:upper} and \eqref{eq:lower}, i.e. a function of column $i$ and $j$ in $P$ and row $i$ and $j$ in $E$
\begin{align}
\begin{split}
     \underline{\Lambda}_{ij} &= \underline{f}(P,E,i,j) := \max_k \left\{ \frac{P_{ki}}{P_{ki} - P_{kj}} \middle| P_{kj} > P_{ki} \right\} \leq 0 \\
    \overline{ \Lambda}_{ij}  &= \overline{f}(P,E,i,j) :=,\min_g \left\{ \frac{E_{jg}}{E_{ig} + E_{jg}}   \middle| E_{ig} + E_{jg} > 0 \right\} \geq 0
\end{split}
\label{eq:range}
\end{align}
such that $\Lambda_{ij} = f(P,E,i,j) = [\underline{\Lambda}_{ij},\overline{ \Lambda}_{ij}]$.

\subsection{Sampling a value $\lambda$ in $\Lambda_{ij}$}
The most simple way to choose $\lambda$ is uniformly at random in the interval $\Lambda_{ij}$, but this choice often gives a slow convergence. The endpoints of the interval often lead to larger changes in $P$ and $E$ and are therefore more favorable. The choice of $\lambda \in \Lambda_{ij}$ is chosen as a shifted symmetric beta distribution with equal shape parameters denoted by $\beta$. Setting $\beta = 1$ results in a uniform choice and setting $\beta < 1$ gives a higher probability at the endpoints. The different choices of $\beta$ are further described in Section \ref{sec:beta} but for now we fix $\beta = 0.5$. Sampling $\lambda \in \Lambda_{ij}$ is made by sampling $x$ from a regular beta distribution with the shape parameters equals to $\beta$ and then setting $\lambda = x \cdot \overline{\Lambda}_{ij} +  (1-x) \cdot \underline{\Lambda}_{ij}$. After sampling a $\lambda \in \Lambda_{ij}$, the matrices $P$ and $E$ are updated to $P_{new} = PA_{ij}(\lambda)$ and $E_{new} = A^{-1}_{ij}(\lambda)E$. In one iteration, this is done sequentially for $i = 1, \dots, N$, where there is chosen a random $j \neq i$ to mix with for each $i$. 

\subsection{Defining the size of the SFS and the stopping criteria}
After each iteration the values of $P$ and $E$ are saved, so after $\mathcal{S}$ iterations we have the following samples from the SFS for both $P$ and $E$:
\begin{align*}
    \textbf{P}^{\mathcal{S}} &=  \{ P^0, P^1, \dots, P^{\mathcal{S}} \} \\
    \textbf{E}^{\mathcal{S}} &= \{ E^0, E^1, \dots, E^{\mathcal{S}} \}.
\end{align*}
This means every entry in $P$ and $E$ have $\mathcal{S}$ samples besides the initial solution $(P^0,E^0)$. We often observe that for some entries all samples are equivalent and for others they spread across an interval. Each new sample is created as an affine combination of the previous sample including itself, which means all the values between two samples in the interval will be feasible as well. The SFS for each entry is therefore defined as the interval between the minimum and the maximum value of these $\mathcal{S}$ samples. Similarly to the polygon inflation algorithm \cite{Sawall2013}, our algorithm also assumes that the SFS consists of connected sets as the samples are affine combinations of one another. This assumption has not appeared to restrict the SFS for any of the data sets we have used, but in theory the solution could have a larger SFS than what is found by our algorithm. We define the size of $\textbf{P}^{\mathcal{S}}$ as the average change of each entry across the full sample:
\begin{equation}
    \text{avg} \langle \textbf{P}^{\mathcal{S}} \rangle = \frac{1}{K \cdot N} \sum_{n = 1}^N \sum_{k = 1}^K  \left\{ \max_{s=0,1,\dots ,\mathcal{S}} P_{kn}^s - \min_{s=0,1,\dots ,\mathcal{S}} P_{kn}^s  \right\}
    \label{eq:avgP}
\end{equation}
To assure that the algorithm stops at the right time, the size of the SFS is calculated after each $\mathcal{T}$ iterations, which we have chosen to fix at $\mathcal{T} = 1000$. The algorithm is stopped when the size of the SFS has changed less than $\epsilon$ within these $\mathcal{T}$ iterations
\begin{equation*}
     \text{avg} \langle \textbf{P}^{\mathcal{S}} \rangle- \text{avg} \langle \textbf{P}^{\mathcal{S} - \mathcal{T} } \rangle < \epsilon,
\end{equation*}
where we set $\epsilon = 10^{-10}$. This assures that our algorithm continues until convergence and stops when no more changes are available. One could add a similar stopping criteria for $\textbf{E}^{\mathcal{S}}$, but in our experiments we found the proposed criteria satisfactory to find the whole SFS for both $P$ and $E$. The sampling algorithm is summarized in Algorithm 1.

\begin{algorithm}
\SetAlgoLined
 Initialize $P^0 = \hat{P}$ and $E^0 = \hat{E}$ \\
 \For{ $s = 1,2, \dots $ }{
 $P_1 \leftarrow P^{s-1}$ \\
 $E_1 \leftarrow E^{s-1}$ \\ 
         \For{ $i = 1, \dots, N$ }{
            $j \leftarrow \text{random element of } \{1,\dots, N\} \backslash \{i\}$ \\
            $\Lambda_{ij} = f(P_i,E_i,i,j)$ \\
            $x \leftarrow \text{Beta}(\beta,\beta)$ \\
            $ \lambda =  x \cdot \overline{\Lambda}_{ij}  +  (1-x) \cdot \underline{\Lambda}_{ij} $ \\
            $P_{i+1} = P_i A_{ij}(\lambda)$ \\
            $E_{i+1} = A_{ij}(-\frac{\lambda}{1-\lambda})E_i $ \\
   }
   $P^s = P_N$ \\
   $E^s = E_N$ \\
   \textbf{stop if} $\text{avg} \langle \textbf{P}^{s} \rangle-\text{avg} \langle \textbf{P}^{s - \mathcal{T}} \rangle< \epsilon$
 }
\caption{Finding the SFS given $(\hat{P}, \hat{E})$  }
\end{algorithm}

\section{Applications and comparison with polygon inflation algorithm}
In this section results from the sampling algorithm are compared with the polygon inflation algorithm. The comparison is made for both the Breast and Lung A. cancer data, when assuming three mutational processes. The results from our sampling algorithm are illustrated in Figure \ref{fig:BRCAsfs} and \ref{fig:LUNGsfs} in the representation defined in \eqref{eq:setPE}. Before making the comparison we have to define another representation of the SFS, that is more commonly used in chemometrics \cite{Borgen1985,Gemper1999,HENRY1990,Sawall2017,Sawall2013} and in particular for the polygon inflation algorithm. After introducing this representation, the assumption of three mutational processes for the comparison is more easily justified.

\subsection{Singular value decomposition to represent the SFS}
In chemometrics it is common to define the SFS relative to the Singular Value Decomposition (SVD). Note, any matrix $PE \in \mathbb{R}_+^{K \times G}$ of rank $N$ can be decomposed into 
\begin{equation*}
    PE = U \Sigma V'
\end{equation*}
where $U \in \mathbb{R}^{K \times N}$ and $V \in \mathbb{R}^{G \times N}$ are the $N$ eigenvectors for $PE (PE)'$ and $(PE)' PE$, respectively, and $\Sigma \in \mathbb{R}^{N \times N}$ is the diagonal matrix of singular values. The factorization into $U$ and $\Sigma V'$ consist of both positive and negative entries, but can be transformed by a matrix $T \in \mathbb{R}^{N \times N}$ such that  $U  T \geq 0$ and $T^{-1} \Sigma V' \geq 0$ \cite{factor2001}. To remove the scaling ambiguities of the factorization, the columns of $T$ are normalised and defined by
\begin{equation}
    T = \left( \begin{array}{c|ccc}
1 &  1 & \cdots &   1 \\ \hline
\alpha_1 & w_{1,1} & \cdots & w_{1,N-1}\\
\vdots & \vdots & \ddots & \vdots \\
\alpha_{N-1} & w_{N-1,1} & \cdots & w_{N-1,N-1}
\end{array} \right) = \begin{pmatrix}
1 & e'\\
\alpha & W 
\end{pmatrix}
\label{eq:Tmat}
\end{equation}
such that $\alpha = (\alpha_1, \dots, \alpha_{N-1})' \in \mathbb{R}^{(N-1)}$, $W  \in \mathbb{R}^{(N-1) \times (N-1)}$ and $e' = (1, \dots, 1) \in \mathbb{R}^{(N-1)}$. The scaling of the columns are made such that the first entry of each column is 1. Here, the SFS is defined by
\begin{align}
    \begin{split}
    \mathcal{M}_{SVD}(P) &= \left\{\alpha \in \mathbb{R}^{(N-1)}  \ \middle| \ \exists W \in \mathbb{R}^{(N-1) \times (N-1)} :  U T \geq 0 \text{ and } T^{-1} \Sigma V' \geq 0 \right\} \\
       \mathcal{M}_{SVD}(E) &= \left\{\alpha \in \mathbb{R}^{(N-1)} \ \middle| \ \exists W \in \mathbb{R}^{(N-1) \times (N-1)} :  U (T')^{-1} \geq 0 \text{ and } T' \Sigma V' \geq 0 \right\} 
    \end{split}
    \label{eq:setalpha}
\end{align}
which is similar to the definition in \cite{Sawall2017,Sawall2013}, where a detailed description of this construction can be seen. As the set in \eqref{eq:setalpha} is defined in terms of $\alpha \in \mathbb{R}^{(N-1)}$, we choose to assume $N = 3$ such that the SFS can be visualized in 2D. 
A big advantage of $\mathcal{M}_{SVD}$ is that it removes the problem of reordering columns and rows as it only focuses on possible transformations of the SVD of $PE$ and not the actual values of $P$ and $E$. 
The connection between the definition in \eqref{eq:setalpha} and the one from \eqref{eq:setPE} is mostly different normalizations of $P$ and $E$. Constructing an invertible $T$ with columns consisting of different $\alpha \in  \mathcal{M}_{SVD}(P)$ gives
\begin{equation*}
    PE =  U \Sigma V' =  \underbrace{U T D_1^{-1}}_{\tilde{P}} \underbrace{D_1 T^{-1} \Sigma V'}_{\tilde{E}}
\end{equation*}
where $\tilde{P} \in \mathcal{M}(P)$, $\tilde{E} \in \mathcal{M}(E)$ and $D_1 = \text{diag}(e'UT) \in \mathbb{R}^{N \times N}$. The diagonal matrix $D_1$ scales the matrices such that $\tilde{P}$ has columns summing to one. Given $\tilde{P} \in \mathcal{M}(P)$, the values of $\alpha$ can be found in the columns of
\begin{equation*}
    T = U' \tilde{P} D_2^{-1}
\end{equation*}
as the eigenvectors in $U$ are orthonormal and $D_2 = \text{diag}(e_1 U' \tilde{P}) \in \mathbb{R}^{N \times N}$, where $e_1 = (1,0,\dots,0)$. The diagonal matrix $D_2$ assures that the first row of $T$ consists of ones as in \eqref{eq:Tmat}. 

The set in \eqref{eq:setPE} is easier to interpret as this visually shows the direct possible changes on $P$ and $E$, but the one in \eqref{eq:setalpha} gives a more simple visualization of the SFS and is better for comparison to the polygon inflation algorithm.

\subsection{Polygon inflation algorithm} 
The polygon inflation algorithm is an algorithm used to approximate the SFS. The algorithm is introduced in \cite{Sawall2013} and a further analysis and proofs are made in \cite{Sawall2017}. The algorithm assumes the set $\mathcal{M}_{SVD}$ for both $P$ and $E$, at most, consists of $N$ separated whole free subsets. In the examples in Figure \ref{fig:SVD} they all have three separated whole free subsets, such that all the points within the polygons are feasible solutions. This assumption is also made for the sampling algorithm, as the SFS are always connected and cannot have two different feasible intervals for one entry. 
The idea of the polygon inflation algorithm is to approximate the subsets of the SFS in terms of the representation in \eqref{eq:setalpha} by inflating an initial solution in $\mathcal{M}_{SVD}$ to the boundaries which creates a polygon. The algorithm continues to inflate the edges of the polygon with new vertices on the boundary of the SFS until the area stops increasing. 
In the package FAC-PACK (\url{http://www.math.uni-rostock.de/facpack/}) they have applied the algorithm for rank three and four. The results from the polygon inflation algorithm and sampling algorithm with rank three are seen in Figure \ref{fig:SVD}, where the grey and black dots illustrate the samples from the sampling algorithm and the colored polygons are the SFS found from the polygon inflation algorithm. 

\begin{figure}[H]
	\centering
	\subfloat[Breast cancer]{\includegraphics[width = 0.8\textwidth]{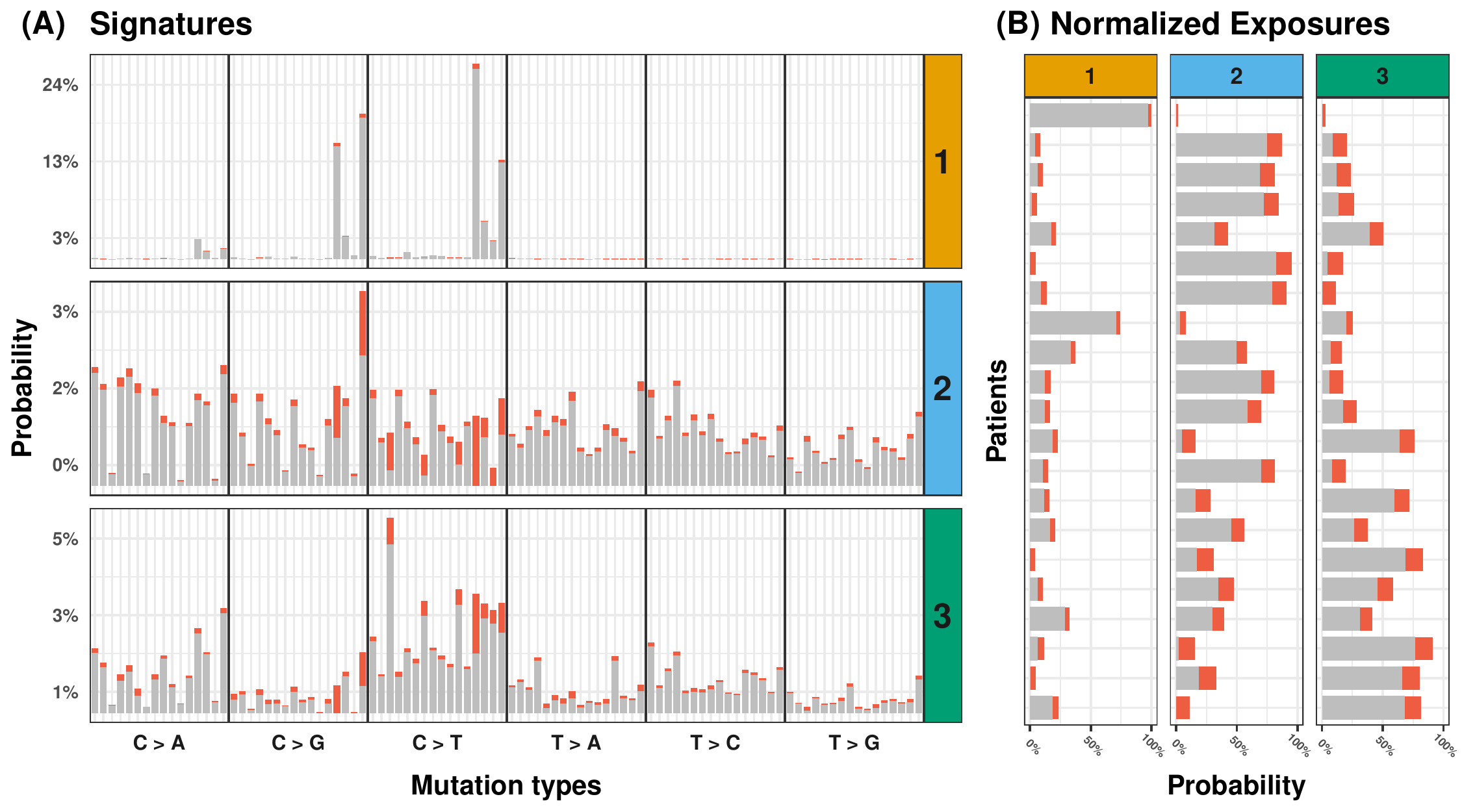} \label{fig:BRCAsfs}}
	\vspace{1cm}
	\subfloat[Lung A. cancer]{\includegraphics[width = 0.8\textwidth]{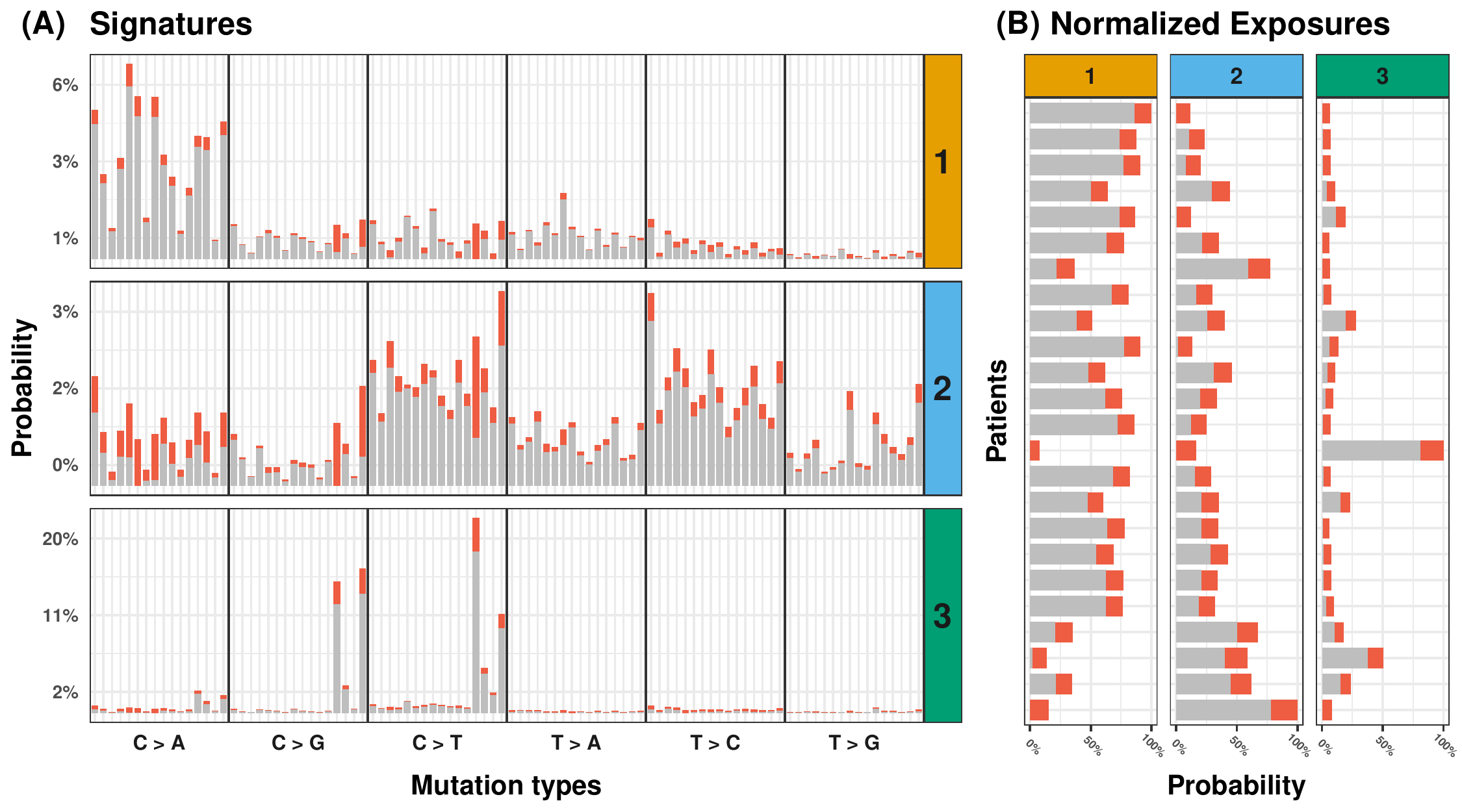}  \label{fig:LUNGsfs}}
	\caption{The signatures and exposures for the Breast and Lung A. cancer data, when assuming three mutational processes. In \textbf{(A)} the signatures are depicted, with the SFS marked in red from the minimum feasible value to the maximum feasible value. In \textbf{(B)} the normalised exposures are depicted and the SFS is again marked in red. The signatures are arranged according to their total exposure i.e. $\sum_{g=1}^G E_{ng}$ in decreasing order. }
	\label{fig:SFSN3}
\end{figure}

\begin{figure}[H]
	\centering
	\begin{center}
		\textbf{Breast Cancer}
	\end{center}
	\vspace{0.2cm}
	\subfloat[$\mathcal{M}_{SVD}(P)$]{
		\includegraphics[height = 6cm, width = 6cm]{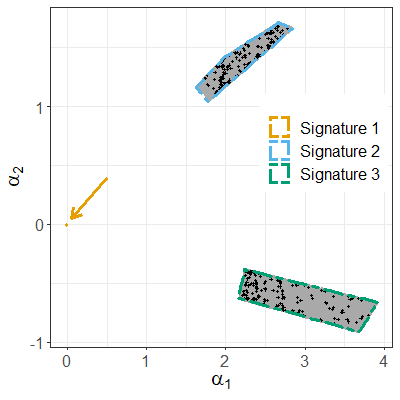}
	} \qquad
	\subfloat[$\mathcal{M}_{SVD}(E)$]{\includegraphics[height = 6cm, width = 6cm]{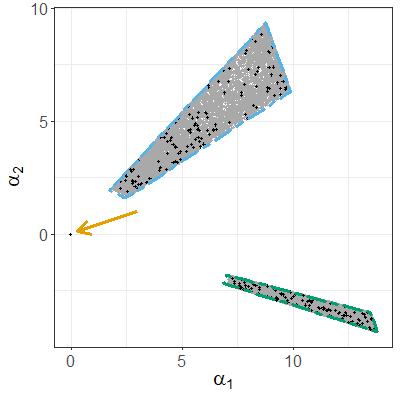} } 
	\vspace{1cm}
	\begin{center}
		\textbf{Lung A. Cancer}
	\end{center}
	\vspace{0.2cm}
	\subfloat[$\mathcal{M}_{SVD}(P)$]{
		\includegraphics[height = 6cm, width = 6cm]{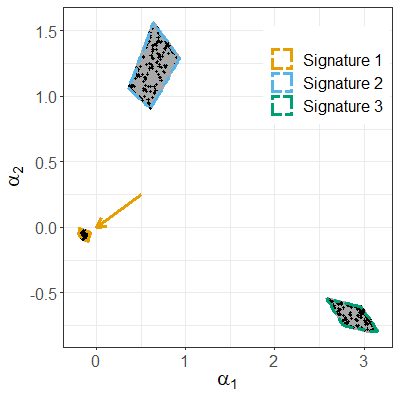}
	} \qquad
	\subfloat[$\mathcal{M}_{SVD}(E)$]{\includegraphics[height = 6cm, width = 6cm]{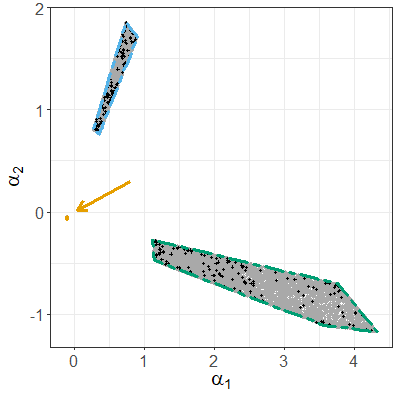} } 
	\caption{The SFS for the Breast and Lung A. cancer data in the $\mathcal{M}_{SVD}$ representation, where the grey dots show the results of all the iterations from the sampling algorithm and the black dots highlight the results from the last 500 samples. The colored polygons are the SFS found from the polygon inflation algorithm. Coloring of the three signatures is in accordance with Figure \ref{fig:SFSN3}. }
	\label{fig:SVD}
\end{figure}
\newpage
The colored lines of the different areas in Figure \ref{fig:SVD} correspond to the signature with the same strip colors in Figure \ref{fig:SFSN3}. The sampling algorithm proposed here clearly gives similar results as the polygon inflation algorithm, which has been the case for all examples we have tried. The SFS for $E$ is also completely found, even though the stopping criteria and sampling is made in relation to the signature matrix $P$. The matrix used as a reference is therefore unimportant in terms of finding the SFS for both $P$ and $E$.

Curious observation is that the size of the areas in Figure \ref{fig:SVD} are hard to directly transfer to the size of the changes in different entries in Figure \ref{fig:SFSN3}. Though, the representation in Figure \ref{fig:SVD} gives a more simple visualization of the SFS. The advantages of our sampling algorithm compared to the polygon inflation algorithm are that it is easier to implement and can scale to an arbitrary dimension of $N$, as shown in Figure \ref{fig:tuning}. This is especially important for cancer genomics where the number of mutational processes varies a lot depending on e.g. the number of samples and number of cancer types. Potentially, this could also be advantageous in other fields where NMF is applied with higher rank of $N$.

\section{Further analysis of the sampling algorithm for arbitrary rank}
Here we will discuss the choice of the parameter $\beta$ and the running time of the sampling algorithm. The sampling algorithm is applied to an assumed rank between 2 and 10, which makes it possible also to comment on how the size of the SFS and computation time is influenced by an increase in the rank. 

\subsection{Influence of tuning parameter $\beta$ and rank $N$ } \label{sec:beta}

\begin{figure}[H]
	\centering
	\subfloat[$\beta = 0.1$]{\includegraphics[width = 4.5cm, height = 4.5cm]{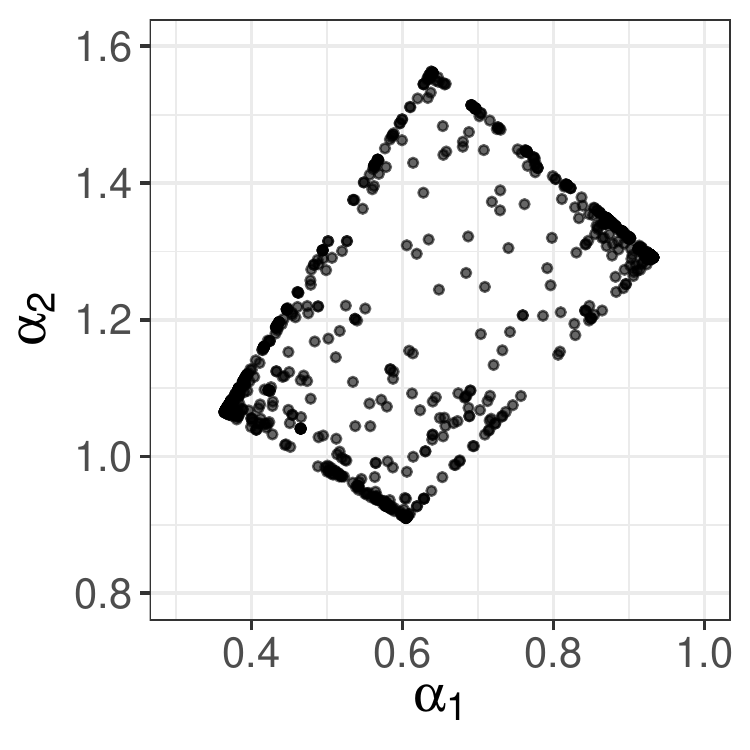}} \qquad 
	\subfloat[$\beta = 0.5$]{\includegraphics[width = 4.5cm, height = 4.5cm]{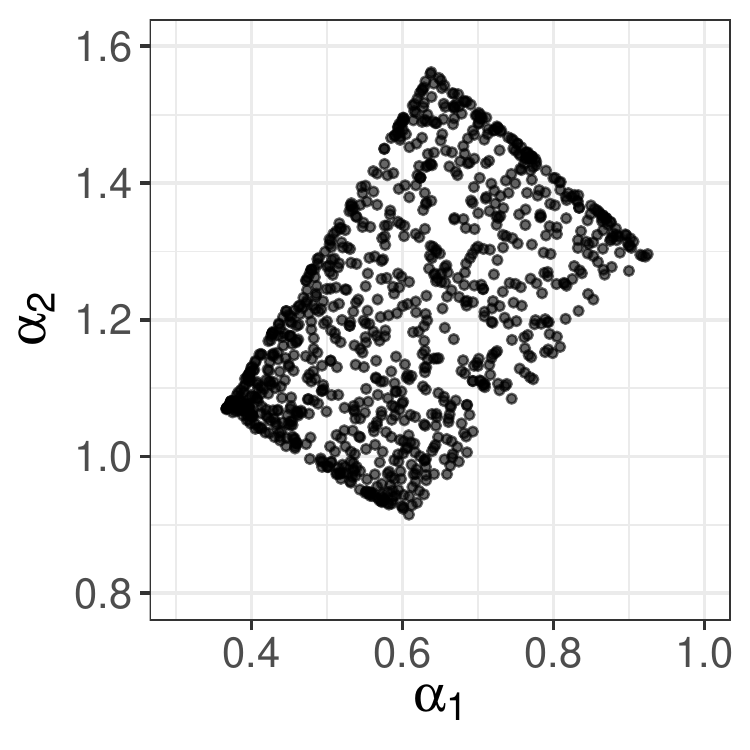}} \qquad 
	\subfloat[$\beta = 1$]{\includegraphics[width = 4.5cm, height = 4.5cm]{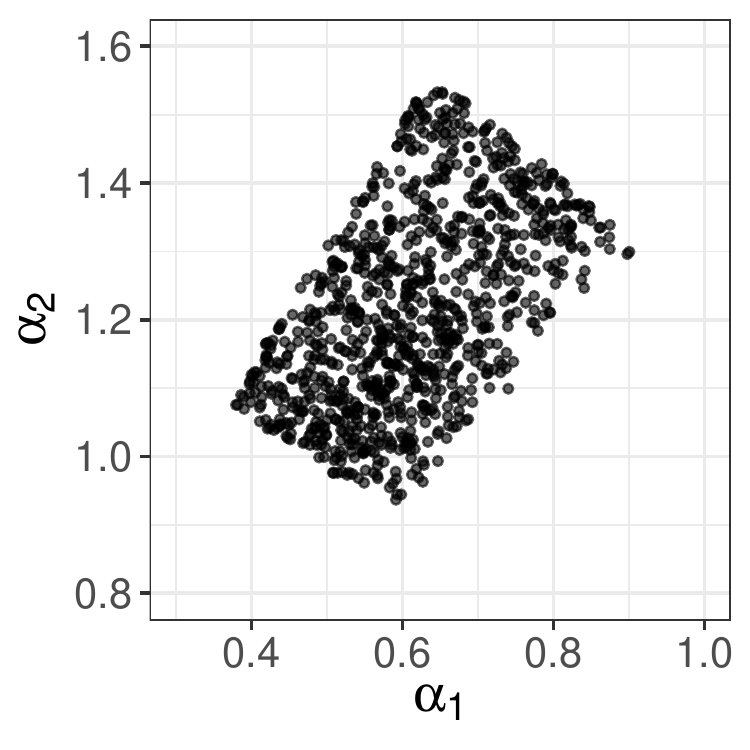}}
	\caption{Sampling 1000 points in the SFS $\mathcal{M}_{SVD}(P)$ for different choices of $\beta$. Here, the results are illustrated on the second signature of Lung A. cancer in Figure \ref{fig:SVD}.}
	\label{fig:sample}
\end{figure}

The influence of the parameter $\beta$ is found by running the sampling algorithm multiple times for $\beta \in \{0.1,0.5,1\}$ on the Breast and Lung A. cancer data. We test for $\beta = 1$ which is equivalent to choosing $\lambda$ uniformly at random and also for $\beta = 0.1$ that mainly samples at the endpoints. At last we also include $\beta = 0.5$, that is something in-between the two cases. An illustration of 1000 samples from the algorithm with $\beta \in \{0.1,0.5,1\}$ is seen in Figure \ref{fig:sample}. Visually we observe from Figure \ref{fig:sample} that $\beta = 0.5$ strikes a good balance between sampling many points close to the edges and at the same time covering the full region of the SFS. 

\begin{figure}[H]
	\centering
	\includegraphics[width = 0.7\textwidth]{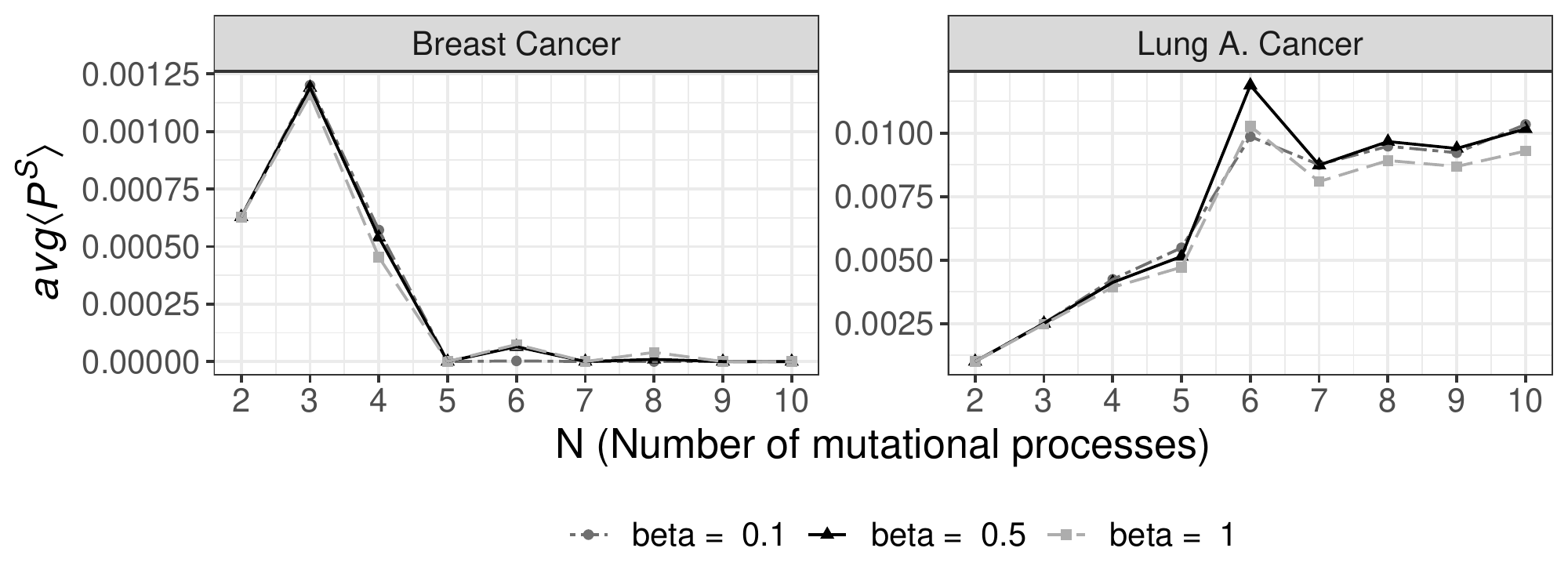} 
	\caption{The size of the SFS for different number of assumed mutational processes i.e. $N$. Each point is based on the results of running the sampling algorithm 50 times for each choice of $\beta$ and $N$. The error bars show the percentile $0.25$ and $0.75$ of the 50 different runs. }
	\label{fig:tuning}
\end{figure}

In Figure \ref{fig:tuning}, the size of the SFS, $\text{avg} \langle \textbf{P}^{\mathcal{S}} \rangle$ is depicted as a function of $N$ i.e. the number of assumed mutational processes. The size of the SFS for the Breast cancer data is close to zero for five mutational processes and above. Even below five mutational processes the average variations of the entries seem to be fairly small.

On the contrary the size of the SFS for the Lung A. cancer is increasing with the number of signatures. The increase of the SFS when $N$ is equal to five instead of three can be seen by comparing Figure \ref{fig:LACAsfs} and Figure \ref{fig:LUNGsfs}, where there are more red areas with $N$ equal to five than with three. Here, it is clear that an increase in the number of mutational processes does not necessarily decrease the size of the SFS. The values $\text{avg} \langle \textbf{P}^{\mathcal{S}} \rangle$ on the y-axis can appear small, but remember from the definition in \eqref{eq:avgP} that we divide our sum by $K \cdot N$.

The choice of $\beta$ seem to have an influence on both the stability and final size of $\text{avg} \langle \textbf{P}^{\mathcal{S}} \rangle$. For five mutational processes and below, the smallest $\beta = 0.1$ performs best, and for more than five mutational processes $\beta = 0.5$ is preferable. Sampling $\lambda \in \Lambda_{ij}$ uniformly i.e. $\beta = 1$ separates the sample equally across the area, which gives a very low probability of sampling in the endpoints and thereby gives a smaller size of the SFS, which are also seen in Figure \ref{fig:tuning}. On the contrary, sampling $\lambda \in \Lambda_{ij}$ with too much weight on the endpoints i.e. $\beta = 0.1$ excludes sampling of points in the center. This works well for a small number of mutational processes, but when the number of mutational processes increases and they can be mixed more across each other it can be hard to reach certain areas of the SFS that requires sampling from the center. We therefore chose to take the middle way with $\beta = 0.5$.

\subsection{Running time}
\begin{figure}[t]
	\centering
	\includegraphics[width = 0.8\textwidth]{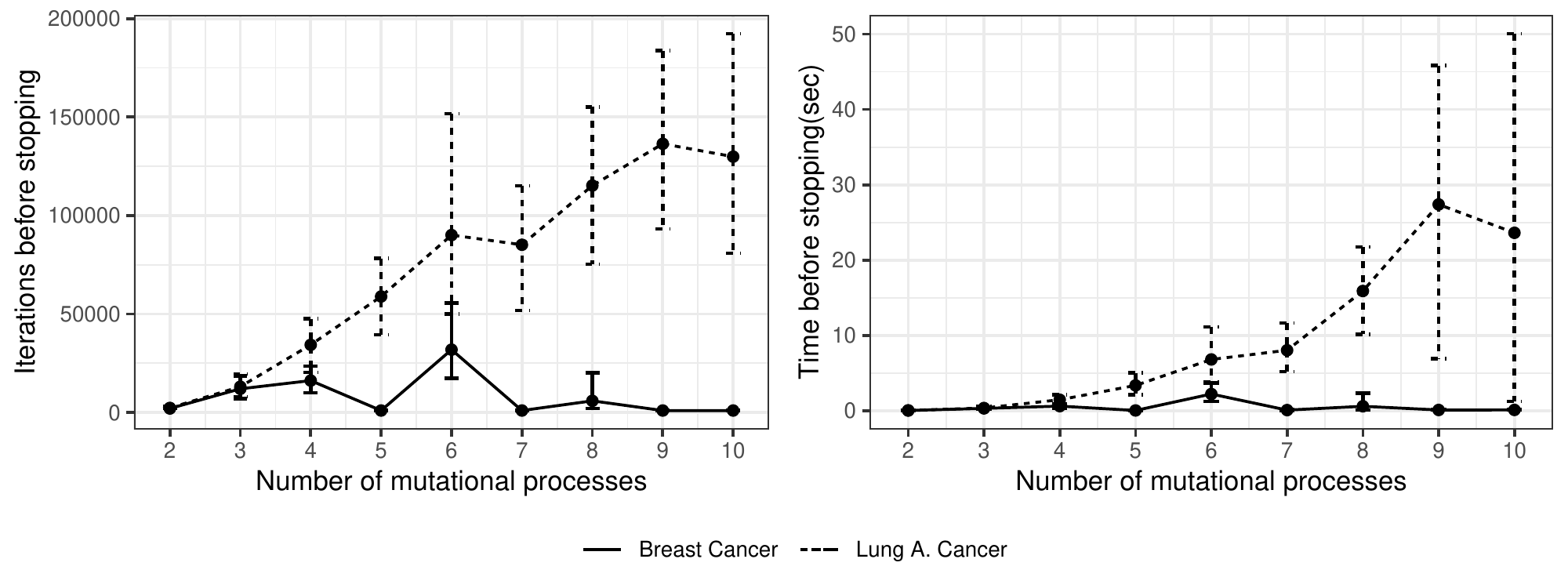}
	\caption{The number of iterations and time needed to compute the SFS for different number of mutational processes, where the algorithm was repeated 50 times. In the figures above, the average is reported together with the quantiles 0.05 and 0.95.}
	\label{fig:complex}
\end{figure}

Our sampling algorithm is not only simple to implement, but is also very fast even for higher ranks of the factorization. In Figure \ref{fig:complex}, the average number of iterations and computation time (using a regular laptop with Intel Core i7-8565U CPU @ 1.80 GHz and 16GB RAM) is shown on the two cancer data sets for different sizes of $N$, where the algorithm was repeated 50 times for each choice of $N$. In general the computation time is highly influenced by the size of the SFS and the rank $N$.

\section{Identification problem due to initialisation and noise compared with the SFS}
The identification of $P$ and $E$ is not only influenced by the size of SFS, but will also be influenced by noise in data and the initialization of the algorithm by Lee and Seung \cite{lee2001}. Here, we will cover the influence of these aspects and compare it to the SFS. In cancer genomics the most natural assumption is that each mutational count is drawn from a Poisson distribution
\begin{equation}
    M_{kg} \sim \text{Pois} \big( (PE)_{kg} \big)
    \label{eq:pois}
\end{equation}
Under this assumption, a natural way to estimate $P$ and $E$ is to maximize the log-likelihood function
\begin{align}
    \ell(P,E; M) &= \sum_{k = 1}^K \sum_{g = 1}^G M_{kg} \log \left( (PE)_{kg} \right) -  (PE)_{kg} - \log \left( M_{kg} ! \right) \\
    &= - D(M|PE) + \textit{C}
\end{align}
where 
\begin{equation}
    D(M|PE) = \sum_{k=1}^K \sum_{g=1}^G M_{kg} \log \left( \frac{M_{kg}}{(PE)_{kg}}\right) - M_{kg} + (PE)_{kg}
    \label{eq:obj}
\end{equation}
is the generalised Kullback-Leibler divergence and \textit{C} is a constant only dependent on $M$. Hence, maximizing the likelihood function is equivalent to minimizing the generalised Kullback-Leibler divergence $D(M|PE)$. This is the divergence appearing in Lee and Seung \cite{lee2001}, which is decreased through each update. Notice, the objective function is convex in either $P$ or $E$, but not in both variables together. As they mention in Lee and Seung \cite{lee2001}, the update rules can therefore not ensure a global minimum, but only a local minimum. Several initialization makes it more likely that a global minimum is reached, which is shown in the next section.

\subsection{Initialisation}
\begin{figure}[H]
	\centering
	\subfloat[Effect of more initializations]{
		\begin{tabular}{cc}
			\quad  \large\textbf{Breast cancer} & \quad \large\textbf{Lung A. cancer} \\
			\includegraphics[width = 0.4\textwidth]{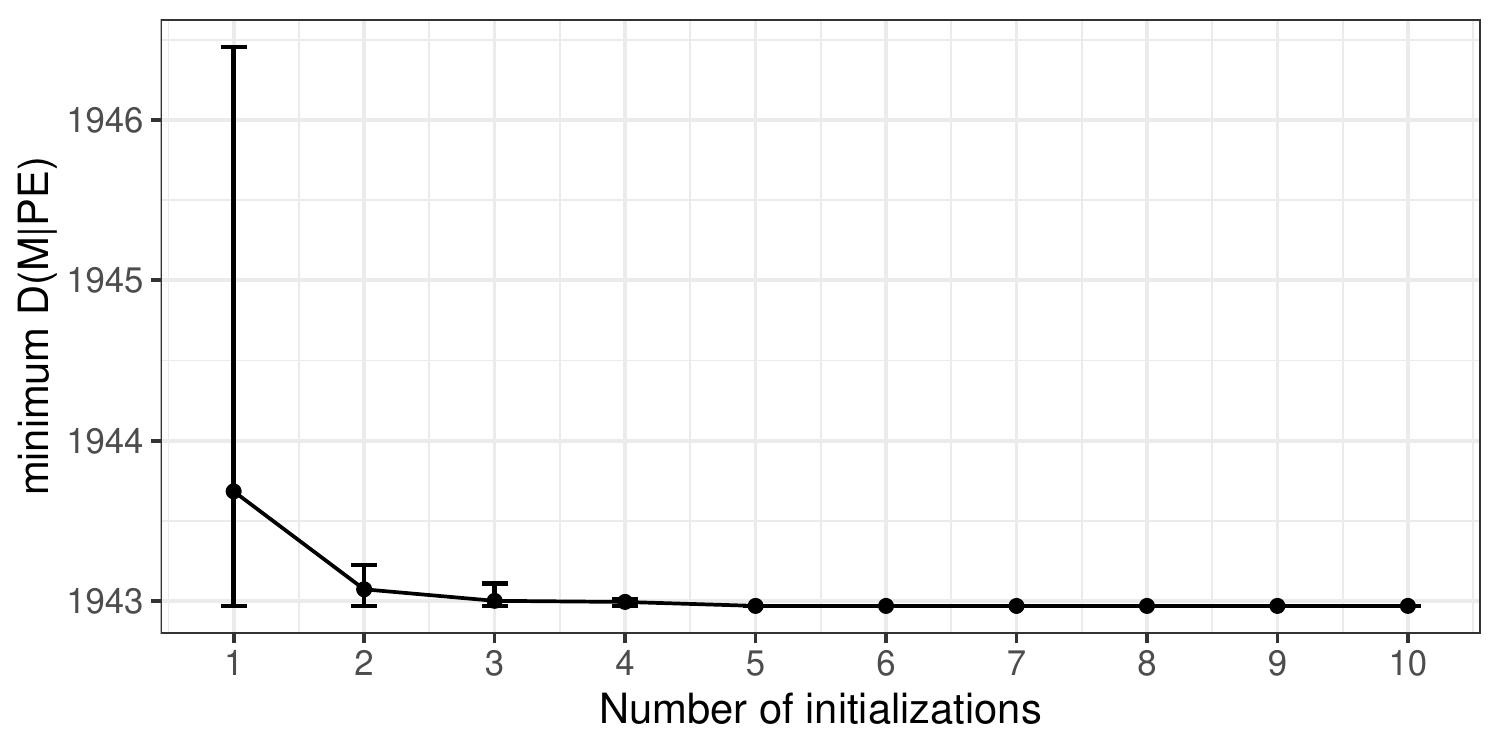}  & \includegraphics[width = 0.4\textwidth]{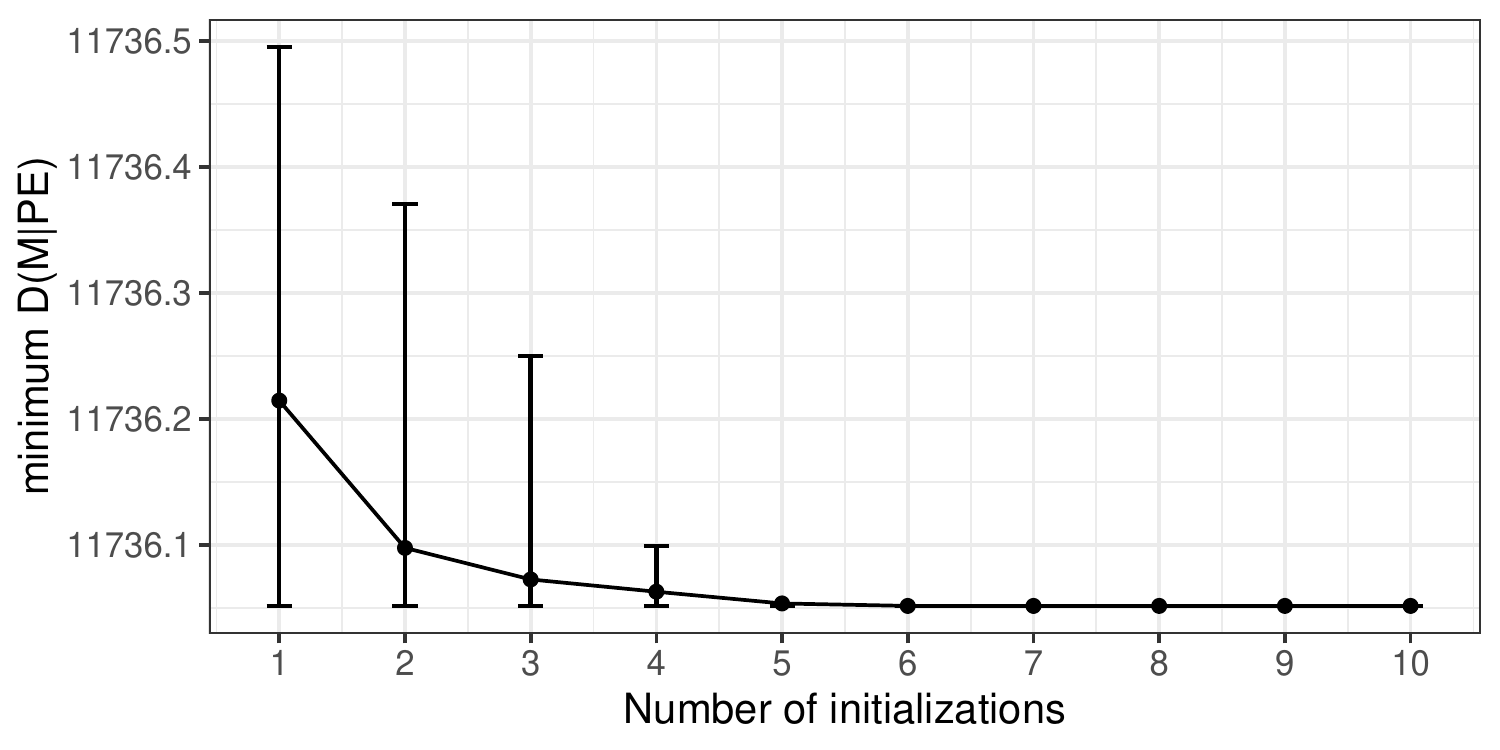}
		\end{tabular}
		\label{fig:inita}
	}\\
	\vspace{1cm}
	\subfloat[Initializations in $\mathcal{M}_{SVD}(P)$]{
		\begin{tabular}{cc}
			\includegraphics[width = 0.35\textwidth]{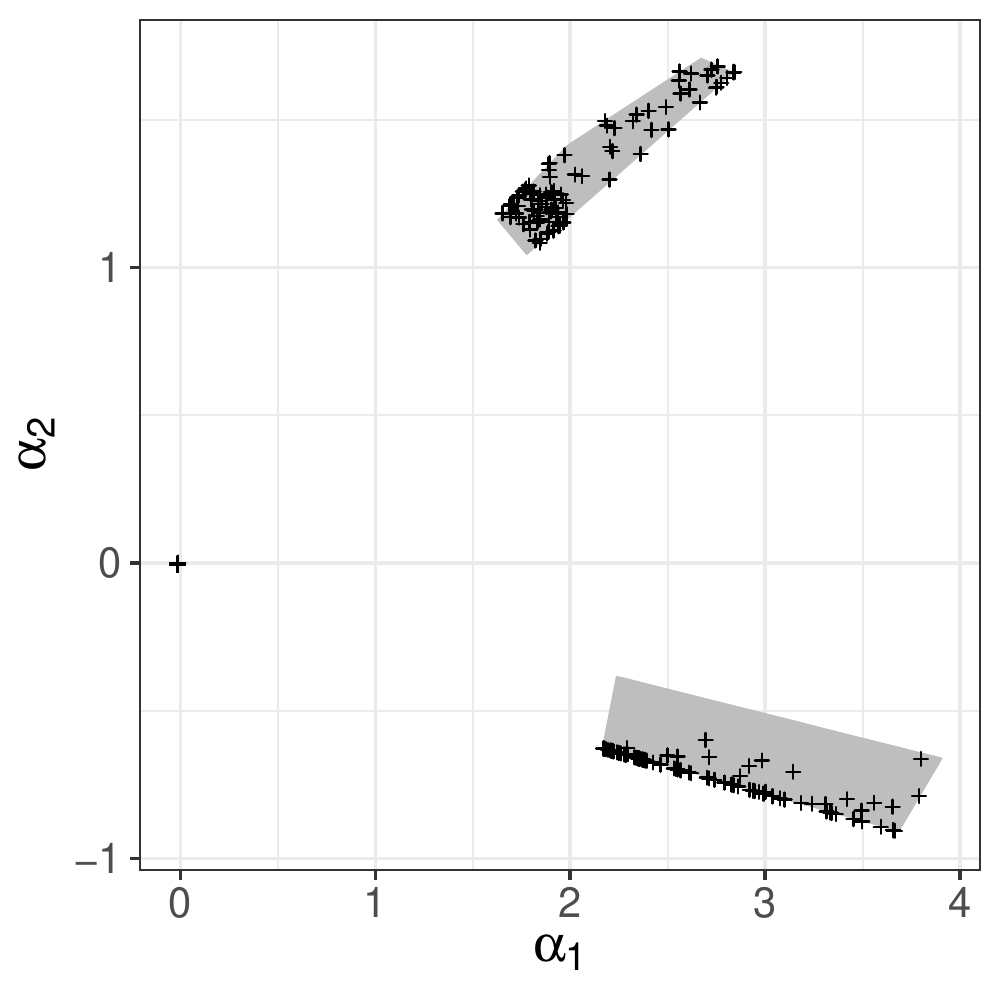} \quad  & \quad \includegraphics[width = 0.35\textwidth]{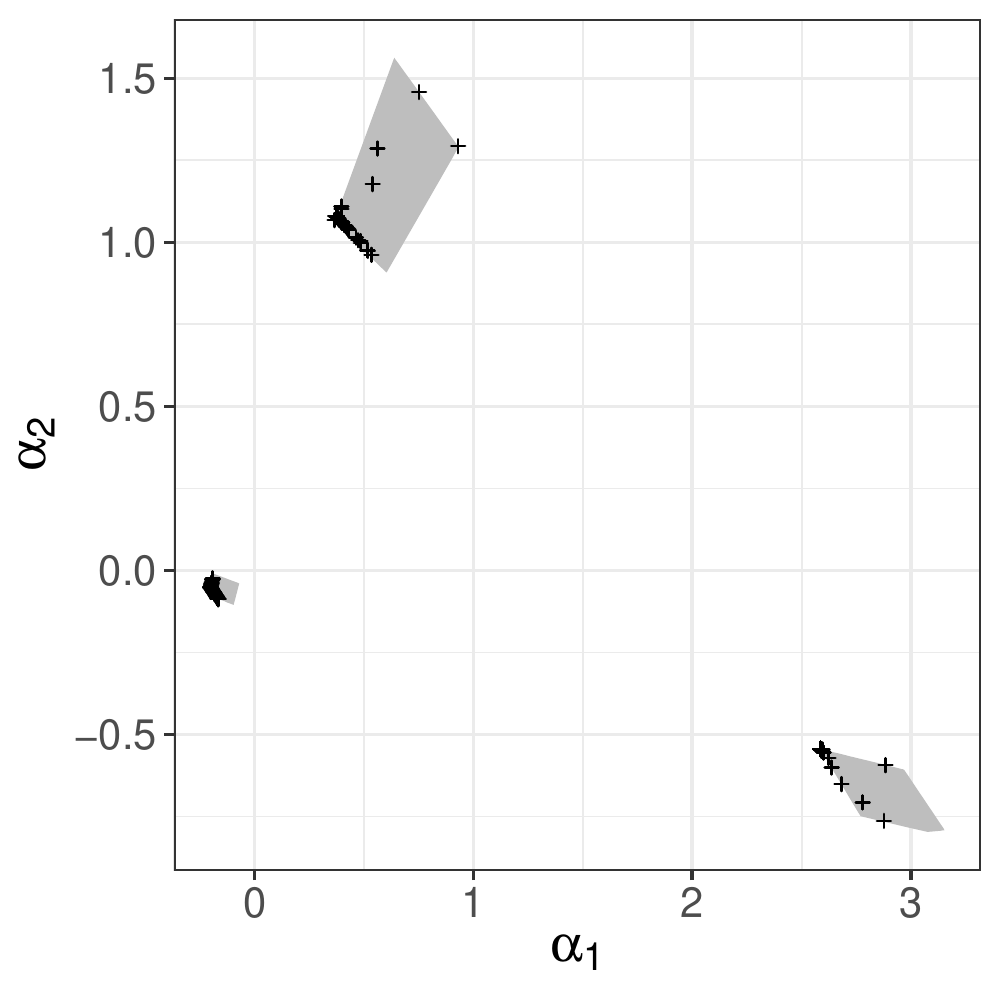}
		\end{tabular}\label{fig:initb}}
	\caption{The influence of initialization. In (a), the average of the minimum $D(M|PE)$ for a certain number of initializations are depicted together with error bars that show the quantiles $0.05$ and $0.95$. The average and quantiles are based on 100 runs. In (b), the results from the 100 different runs with ten initializations are shown in $\mathcal{M}_{SVD}(P)$ relative to the solution found previously in Figure \ref{fig:SVD}(a).}
\end{figure}

Through this paper, we have performed at least five random initializations when finding an NMF solution. The influence of an increase in the number of initializations can be seen in Figure \ref{fig:inita}, which shows how just a few initializations will stabilize the reached minimum to a large extend. As the minimum stops changing we assume to have found the global minimum. Even though the algorithm reaches the same minimum we will still observe variability in the estimates of $P$ and $E$, when they have a variability in the SFS.  

Different random initializations with the same minimum will still give different results in the SFS, which is illustrated in Figure \ref{fig:initb}. Some of the SFS can therefore appear through the initialization, but using initializations to find the SFS would be a very unreliable and time consuming way to recover the SFS. Also, it is clear to observe in Figure \ref{fig:initb} how the iterative procedure often converges to the same areas of the SFS. 

\subsection{Noise in the data}
\begin{figure}[hb]
	\centering
	\subfloat[Random initialization]{\includegraphics[width = 0.4\textwidth]{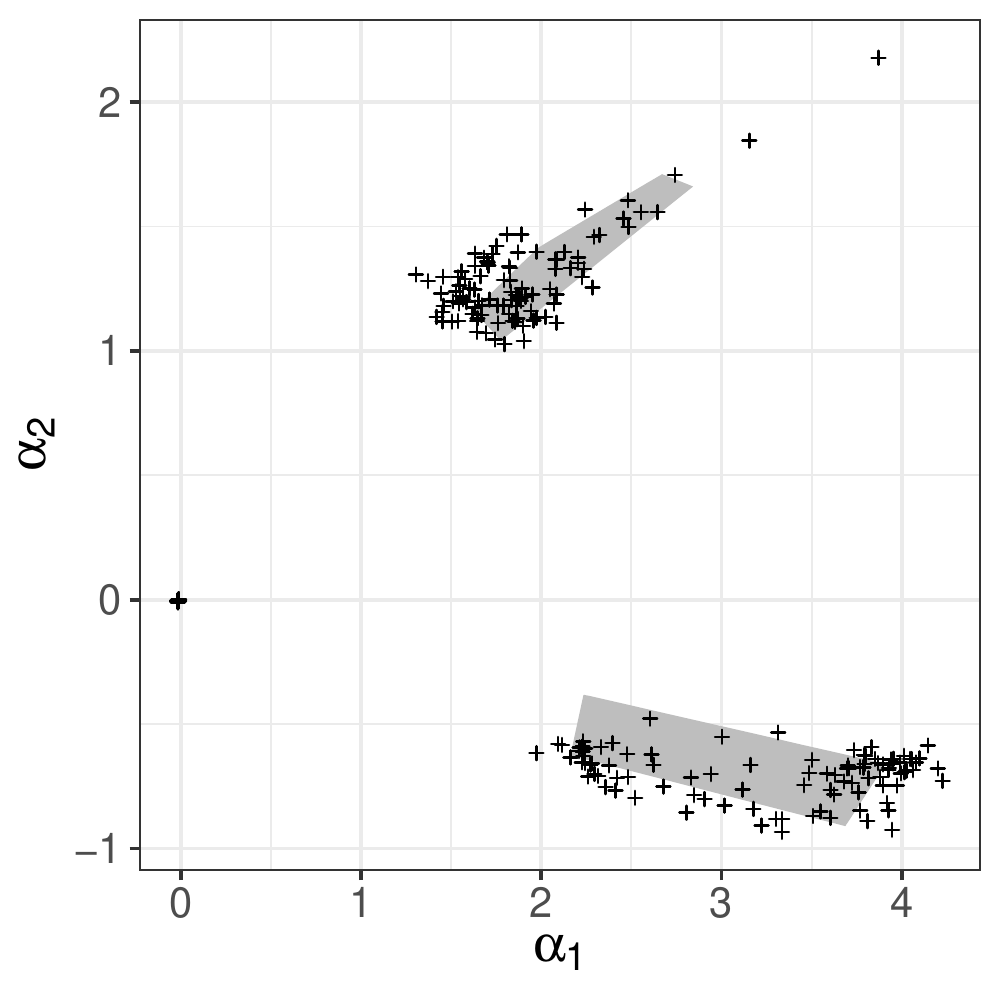}} \quad 
	\quad \subfloat[Same initialization]{\includegraphics[width = 0.4\textwidth]{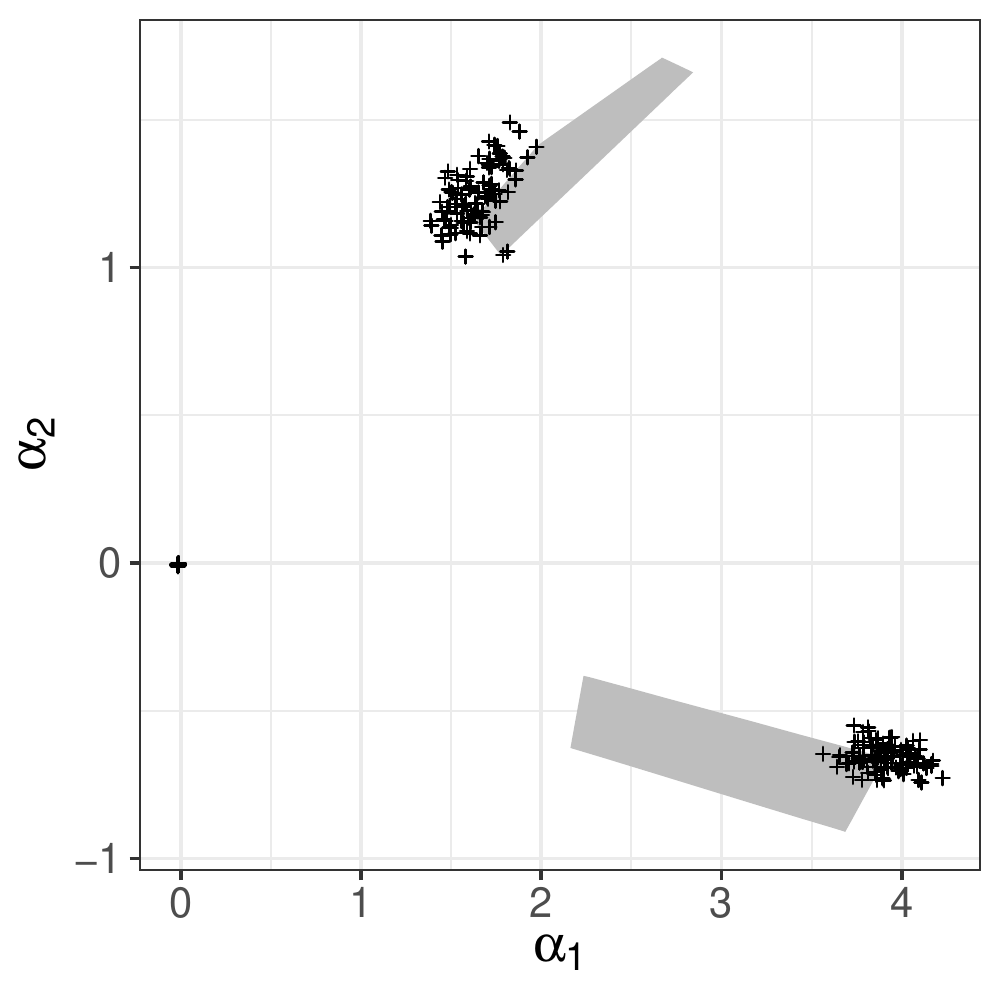}}
	\caption{Visualization of the identification problem due to variance, where (a) shows the results from random initialization and (b) shows the results using the same ten initialization for each sample. Both show the global minimum from 100 bootstrap samples.}
	\label{fig:boot}
\end{figure}

Noise in the data obviously has an influence on the identification of $P$ and $E$, but sometimes the influence can seem higher than it actually is due to variability in the SFS. We will illustrate this on the Breast cancer data assuming three mutational processes, where we have variability in the SFS. Assume we have found a global minimum factorization $\hat{P}\hat{E} \in \mathbb{R}_+^{96 \times 21}$ that approximates the Breast cancer data $M \in \mathbb{N}_0^{96 \times 21}$. The influence of variance is now illustrated through parametric bootstrapping using the assumption in \eqref{eq:pois}. Specifically, we create 100 bootstrap samples 
\[M^{*b} \sim \text{Pois}(\hat{P}\hat{E})\]
for $b = 1, \dots , 100$. The NMF algorithm from Lee and Seung \cite{lee2001} is then run on $(M^{*1},\dots M^{*100})$ for both 10 different random initializations and the same 10 initializations, which is illustrated in Figure \ref{fig:boot}. We clearly see how the appearing variability of the results is influenced by the SFS through random initialization. 

A large variance could therefore appear as a low reproducibility of the signatures, but is actually just a result of the SFS. Remember, having a large SFS does not make the assumed number of mutational processes less correct but is instead a result of the structure of the signatures and exposures. 

\section{Conclusion}
We introduce a novel sampling algorithm that can find the SFS for an arbitrary rank of NMF. The algorithm is based on the simple and well known calculations for two dimensions together with random sampling to explore the possible changes of the initial solution. The algorithm also gives equivalent results when compared to current algorithms, most notably the polygon inflation algorithm. The problem of non-uniqueness for NMF in relation to mutational counts have also been highlighted here. The size of the SFS is shown to depend strongly on the specific data set at hand and the assumed number of mutational processes. Furthermore, the importance of several initializations is enlightened together with the important fact of how a large variance in a signature could potentially be caused by the SFS. 

An important aspect of NMF is to recover the true rank of the factorization. In cancer genomics there have both been introduced methods based on the Bayesian Information Criteria (BIC) \cite{fischer2013,signeR2017} and another method that both includes the variability of the signatures and how well $P$ and $E$ approximates $M$ \cite{alex2013}. The latter is definitely more questionable, as there could appear large variations due to the size of the SFS. An interesting future problem is the influence of having a SFS in the further analysis of the mutational processes. Would the conclusion from one solution in the SFS be very different from another? One method to remove the problem of non-uniqueness of the solution is by decreasing the variability of the signatures through fewer parameters. This was implemented in a simple form in \cite{shiraishi2015}. 

Our sampling algorithm is very simple to implement and fast, which makes it easy to check whether a NMF solution is unique or not. This article has enlightened the problem of non-uniqueness of the NMF and at the same time showed a simple way to check this uncovered problem. It is important to remember that the size of the SFS is not a tool to determine the correct choice of $N$, but instead a way to find all the solutions that give the same global minimum approximation of the data. Variability in the SFS is not the result of a poor factorization, but instead lack of uniqueness of the matrix factorization.

\subsection*{Acknowledgements}
We would like to thank Joachim Beck for his master thesis on this subject, as it gave much inspiration to the development of the sampling algorithm. Furthermore, we would like to thank Dan Ariel Søndergaard and Kenneth Borup for helpful suggestions and corrections to the manuscript.

\nolinenumbers
\bibliography{ArticleRLAH}

\end{document}